\RequirePackage{lineno}

\documentclass[final,5p,twocolumn,preprint,sort&compress]{elsarticle}

\usepackage{graphicx}  
\usepackage{dcolumn}   
\usepackage{bm}        
\usepackage{amssymb}   

\usepackage{mathptmx}
\usepackage{lineno}
\usepackage{hyperref}
\usepackage{xspace}
\usepackage{multirow}
\usepackage{epsfig}
\usepackage{psfrag}
\usepackage{lscape}
\usepackage{comment}

\newcommand{\ttbar}{\ensuremath{t\bar{t}}\xspace}
\newcommand{\etmiss}{\ensuremath{E \kern-0.6em\slash_{\rm T}}\xspace}
\newcommand{\etmissx}{\ensuremath{E \kern-0.6em\slash_{\rm x}}\xspace}
\newcommand{\etmissy}{\ensuremath{E \kern-0.6em\slash_{\rm y}}\xspace}

\newcommand{\pythia}{{\sc pythia}\xspace}

\newcommand{\vecbos}{{\sc vecbos}\xspace}

\newcommand{\dm}{\ensuremath{\Delta m}\xspace}

\newcommand{\ejets}{\ensuremath{e\!+\!{\rm jets}}\xspace}
\newcommand{\mujets}{\ensuremath{\mu\!+\!{\rm jets}}\xspace}

\newcommand{\pevt}{\ensuremath{P_{\rm evt}}\xspace}
\newcommand{\psig}{\ensuremath{P_{\rm sig}}\xspace}
\newcommand{\pbkg}{\ensuremath{P_{\rm bkg}}\xspace}
\newcommand{\kjes}{\ensuremath{k_{\rm JES}}\xspace}
\newcommand{\dstar}{\ensuremath{D^*}\xspace}

\newcommand{\x}{\ensuremath{\vec x}\xspace}
\newcommand{\y}{\ensuremath{\vec y}\xspace}

\newcommand{\eps}{\varepsilon}

\newcommand{\GeV}{\ensuremath{\textnormal{GeV}}\xspace}

\newcommand{\dif}{\ensuremath{{\rm d}}}

\newcommand{\fb}{\ensuremath{{\rm fb}^{-1}}\xspace}

\newcommand{\mt}{\ensuremath{m_t}\xspace}
\newcommand{\mtfit}{\ensuremath{m_t^{\rm fit}}\xspace}
\newcommand{\mtgen}{\ensuremath{m_t^{\rm gen}}\xspace}

\newcommand{\pt}{\ensuremath{p_T}\xspace}

\journal{Nuclear Instruments and Methods in Physics Research A}

\begin{document}



\begin{frontmatter}
\title{Acceleration of matrix element computations for precision measurements\tnoteref{t1}}

\author[goe,hd]{O.~Brandt\corref{cor1}}
\ead{obrandt@fnal.gov}
\author[fnal]{G.~Gutierrez}
\author[fnal]{M.H.L.S.~Wang} 
\author[uic]{Z.~Ye}
\address[goe]{II. Physikalisches Institut, Georg-August-Universit\"at G\"ottingen, G\"ottingen, Germany}
\address[hd]{now at Kirchhoff-Institut f\"ur Physik, Universit\"at Heidelberg, Heidelberg, Germany}
\address[fnal]{Fermi National Accelerator Laboratory, Batavia, Illinois 60510, USA}
\address[uic]{University of Illinois at Chicago, Chicago, Illinois 60607, USA}

\tnotetext[t1]{FNAL report number: Fermilab-Pub-14/416-E.}
\cortext[cor1]{Corresponding author}

\date{October 23, 2014}

\begin{abstract}
The matrix element technique provides a superior statistical sensitivity for precision measurements of important parameters at hadron colliders, such as the mass of the top quark or the cross section for the production of Higgs bosons. The main practical limitation of the technique is its high computational demand. Using the concrete example of the top quark mass, we present two approaches to reduce the computation time of the technique by a factor of 90. First, we utilize low-discrepancy sequences for numerical Monte Carlo integration in conjunction with a dedicated estimator of numerical uncertainty, a novelty in the context of the matrix element technique. Second, we utilize a new approach that factorizes the overall jet energy scale from the matrix element computation, a novelty in the context of top quark mass measurements. The utilization of low-discrepancy sequences is of  particular general interest, as it is universally applicable to Monte Carlo integration, and independent of the computing environment.

\end{abstract}


\end{frontmatter}


\section{Introduction}
The matrix element~(ME) technique~\cite{bib:kondo} is a powerful tool in experimental particle physics, especially at hadron colliders, as it provides a superior statistical sensitivity in the extraction of important parameters of the standard model. This sensitivity is achieved by taking into account the full topological and kinematic information in a given event, and determining the probabilities \psig and \pbkg for observing each event, assuming respective signal and background hypotheses in the respective ME probabilities~$|\mathcal M_{\rm sig}|^2$ and $|\mathcal M_{\rm bkg}|^2$. In the context of searches for new physics, these probabilities can be used to construct the most powerful test statistic $Q\equiv\frac{\psig}{\pbkg}$ according to the Neyman-Pearson lemma~\cite{bib:neyman}. An advantage of the ME technique is that it calculates \psig and \pbkg\ {\em ab initio}, in contrast to multivariate methods. Furthermore, \psig depends directly on the physical parameter of interest in a specific theoretical framework.

The ME technique was first suggested by Kondo~\cite{bib:kondo} and pioneered in the context of experimental particle physics at the Tevatron in the measurement of the mass of the top quark \mt~\cite{bib:mtopnature}, in the determination of the helicity of the $W$ boson~\cite{bib:whel}, as well as for the first evidence for production of single top quarks~\cite{bib:singletopd0,bib:singletopcdf}. Since then, the ME technique has been used in several analyses, for example in searches for the Higgs boson at the Tevatron~\cite{bib:higgscdf} and at the LHC~\cite{bib:higgscms}. Recently, a general framework for the ME technique, named MadWeight~\cite{bib:madweight}, has become available.

Despite its superior statistical sensitivity, the ME technique is not widely applied because of its high computational demand. For example, to perform a previous measurement of \mt using 3.6~\fb\ of integrated luminosity~\cite{bib:mtop36} by the D0 Collaboration, about two million CPU-hours were required on a single core of the 64 bit XEON E5-2620 CPU, with a clock rate of 2~GHz, and a 64 bit computation.
In this manuscript, we present two approaches that were successfully applied to reduce the computational demand of the ME technique by two orders of magnitude.
First, we utilize low-discrepancy sequences  (LDS) for numerical Monte Carlo (MC) integration, in conjunction with a dedicated estimator of the numerical uncertainty, which is a novelty in the context of the ME technique. Second, we factorize the overall jet energy scale (JES) from the ME computation, which was never done before in the context of \mt measurements using an {\it in situ} JES calibration. The use of LDS is generally applicable to MC integration. In particular, this approach is not  hardware-specific, i.e., it can be used on, e.g., a graphical processing unit. 

We present our results using the example of the recent measurement of the top quark mass~\cite{bib:mtop}, the single most precise measurement of this parameter, yielding $\mt=174.98 \pm 0.58\thinspace({\rm stat}) \pm 0.49\thinspace({\rm syst})~\GeV$.  This measurement was performed in lepton$+$jets final states\footnote{The lepton$+$jets final states aim at selecting the $p\bar p\to\ttbar\to W^+bW^-\bar b\to\ell^+\nu bq\bar q'\bar b$ and its charge conjugate process, where $t$ and $b$ denote respectively top and bottom quarks, $W^\pm$ is the $W$ boson, $\ell^\pm$ stands for charged leptons, and $\nu$ represents a neutrino.} with the full sample of $p\bar{p}$ collision data from the Fermilab Tevatron Collider at \mbox{$\sqrt s=1.96~$TeV}, corresponding to $9.7~\fb$ of integrated luminosity. The computational demand arises not so much from the number of events recorded in $p\bar p$ collisions, but rather from number of the simulated MC events which are used for the calibration of the method and for the evaluation of systematic uncertainties. D0's previous measurements of \mt~\cite{bib:mtop36} and of the difference $\dm=m_t-m_{\bar t}$~\cite{bib:dm36}, both using 3.6~\fb of integrated luminosity, were also performed with the ME technique.

This manuscript is structured as follows. We begin with a brief review of our previous implementation of the ME technique for the measurement of \mt~\cite{bib:mtop36} in 3.6~\fb of data. This analysis applies several approaches to reduce the computational demand that potentially have general interest. We follow with a discussion of our latest implementation of the ME technique, which provides further reduction in the computational demand through use of LDS for the MC integration, presented in Sec.~\ref{sec:lds}, and through factorization of the scale factor for jet energies $\kjes$ from the ME computation, discussed in Sec.~\ref{sec:kjes}. Finally, we present in Sec.~\ref{sec:val} the validation of our latest implementation of the ME technique with pseudo-experiments (PE), comprised of MC events fully simulated in the D0 detector, and conclude in Sec.~\ref{sec:conclusion}. The MC simulations are described in Ref.~\cite{bib:mtop}.

\section{Previous implementation of the matrix element technique\label{sec:previous}}

The extraction of \mt with the ME technique 
is performed with a likelihood that uses per-event probability densities (PD) defined by the ME of the processes contributing to the observed events. 
Assuming two non-interfering contributions from \ttbar and $W+{\rm jets}$ production, the per-event PD is given by
\begin{linenomath}
\begin{eqnarray}
\pevt &=& A(\x)[ f\psig(\x; \mt,\kjes)\nonumber\\
      &+& (1-f)\pbkg(\x;\kjes) ]\,,\label{eq:pevt}
\end{eqnarray}
\end{linenomath}
where the observed signal fraction $f$, \mt, and the overall multiplicative factor $\kjes$ adjusting the energies of jets after their default jet energy scale calibration, are parameters to be determined from data. The $\x$ denotes the measured jet and lepton four-momenta, and $A(\x)$ accounts for acceptance and efficiencies. 
The function \psig represents the PD for \ttbar production, and \pbkg refers to the PD for $W+{\rm jets}$ production.

In general, the measured set $\x$ will not be identical to the set of corresponding partonic variables $\y$ because of finite detector resolution and parton hadronization. Their relationship is described by a transfer function $W(\x,\y,\kjes)$. 
The densities \psig and \pbkg are calculated through a convolution of the differential partonic cross section, $\dif\sigma(\y)$, with $W(\x,\y,\kjes)$ for the final-state partons and the PD for the initial-state partons, $f(q_i)$, where the $q_i$ are the momenta of the colliding partons. This is done by integrating over all possible parton states that lead to $\x$:
\begin{linenomath}
\begin{eqnarray}
\psig(\x;\mt,\kjes) &=& \frac1{\sigma_{\ttbar,\rm obs}(\mt,\kjes)}\int\sum\dif\sigma(\y,\mt)\dif \vec q_1\dif\vec q_2 \nonumber\\
&\times& f(\vec q_1)f(\vec q_2) W(\x,\y;\kjes)\,.
\label{eq:psig}
\end{eqnarray}
\end{linenomath}
The sum extends over all possible flavor combinations of the initial-state partons. The longitudinal-momentum parton density functions (PDF) $f(q_{i,{\rm z}})$ are taken from the CTEQ6L1 set~\cite{bib:cteq}, while the dependencies $f(q_{i,{\rm x}})$, $f(q_{i,{\rm y}})$ on transverse momenta follow those PD obtained from the \pythia simulation~\cite{bib:pythia1,bib:pythia2}. The factor ${\sigma_{\ttbar,\rm obs}(\mt,\kjes)}$, defined as the total cross section for \ttbar production in $p\bar p$ collisions to be observed in the detector,
ensures that $A(\x)\psig$ is normalized to unity. The differential cross section, $\dif\sigma(\y,\mt)$, is calculated using the leading order (LO) ME for the process $q\bar q\to\ttbar$~\cite{bib:melo1,bib:melo2}.

The calculation in Eq.~\ref{eq:psig} at LO involves 24 integration variables associated with the two initial-state partons and the six partons in the final state. The directions of the four jets and the charged lepton in $(\eta,\phi)$ space are well-measured, and are therefore represented by ten $\delta$-functions. After accounting for these $\delta$-functions, and imposing energy-momentum conservation through four additional $\delta$-functions, ten integration variables remain.

The integration in Eq.~\ref{eq:psig} is performed numerically using the MC integration method of Ref.~\cite{bib:mcmethod}. The pseudo-random numbers for the MC integration are generated with {\sc ranlux}~\cite{bib:ranlux} in a $[0,1]^{10}$ hypercube, and then transformed to the ranges of the integration variables. Importance sampling~\cite{bib:numrec} is utilized to reduce the computational demand of the integration. Furthermore, we perform a Jacobian transformation of the nominal ten integration variables to variables where prior information is either known or can be easily obtained. This prior information is then used in importance sampling. The optimized integration variables are: $m_{W^+},~m_{W^-}$, $m_t,~m_{\bar t},~q_{1,\rm x},~q_{1,\rm y},~q_{2,\rm x},~q_{2,\rm y}$, $\rho=E_q/(E_q+E_{\bar q'})$ for the quarks from $W\to q\bar q'$ decay in the LO picture where $E$ represents the particle's energy, and, the energy (curvature) of the electron (muon track)  $\kappa$.

To integrate over $m_t$ and $m_{\bar t}$, random numbers are generated according to expected Breit-Wigner distributions for each given $\mt$ hypothesis. The constraint of $M_W = 80.4$~GeV for the {\it in-situ} JES calibration is imposed by integrating over $W$ boson masses using a Breit-Wigner prior. For the integration over $q_{i,\rm x}$ and $q_{i,\rm y}$, the ME is sampled in transverse momentum $p_T^{q_i}$  according to the distribution predicted in MC simulations, and uniformly in $\phi^{q_i}$. To integrate over $\kappa$, random numbers are generated according to the corresponding part of the transfer function, which is defined as the probability to obtain the measured $\kappa_x$ value, given a value $\kappa_y$ at the parton level.

Importance sampling in ten bins is employed for the integration over $\rho$. The MC integration is performed iteratively with an increasing number of samplings of the integral per iteration, where each iteration uses the probability distribution in $\rho$ from the previous one as input for importance sampling.

There are 24 possible jet-parton assignments that are summed with weights based on their consistency with $b$-tagging information\footnote{We identify jets from $b$ quarks  through the use of a multivariate algorithm, as discussed in Ref.~\cite{bib:mtop}.}. Typically, two and sometimes four or six jet-parton assignments numerically dominate the final result for \psig. To identify them, we perform a pre-integration step, where we calculate $\psig^i$ for each jet-parton assignment $i$, until a relative numerical precision of 10\% is reached, or the integral is sampled $2^{14}=16,384$ times. The numerical precision of those jet-parton assignments with $\psig^i$ within 2\% of the maximal $\psig^i$ value is further refined until the desired precision has been achieved, or the integral is sampled $2^{24}=16,777,216$ times. For all other assignments $\psig^i$ obtained in the pre-integration step is kept.

The differential partonic cross section for \pbkg is calculated similarly to \psig, i.e., applying MC integration and the same transfer function $W(\x,\y;\kjes)$, however using the LO $W+4{\rm~jets}$ ME implemented in \vecbos~\cite{bib:vecbos}. Here, the initial-state partons are all assumed to have no transverse momentum $\pt=0$.

To extract \mt and \kjes, we calculate \psig and \pbkg on a grid in $(\mt,\kjes)$ with spacings of $(1~\GeV,0.01)$. A likelihood function ${\cal L}(\x_1,\x_2,...,\x_N;\mt,\kjes,f)$ is constructed at each grid point from the product of the individual \pevt values for the measured quantities $\x_1,\x_2,...,\x_N$ of the selected events, and the signal fraction $f$ is determined by maximizing $\cal L$ at that grid point. The likelihood function ${\cal L}(\x_1,\x_2,...,\x_N;\mt,\kjes)$ is then projected onto the $\mt$ and $\kjes$ axes by integrating using Simpson's rule~\cite{bib:simpson} over \kjes and \mt, respectively. Best unbiased estimates of \mt and \kjes and their statistical uncertainties are extracted from the mean and standard deviation (SD) of  ${\cal L}(\x_1,\x_2,...,\x_N;\mt)$ and  ${\cal L}(\x_1,\x_2,...,\x_N;\kjes)$.

Further details on the previous implementation of the ME technique can be found in Ref.~\cite{bib:mtop36}.

\section{
Reducing the computation demand of the matrix element technique with low-discrepancy sequences in MC integration
\label{sec:lds}
}

The expected uncertainty of the MC integration method based on classical pseudo-random number sequences, decreases as
\[
\frac1{\sqrt N}\,,{\rm ~for~}N\to\infty\,,
\]
where $N$ is the number of integral samplings, i.e., points in the $[0,1]^d$ unit hypercube of dimension $d$ for which the integrand is evaluated~\cite{bib:numrec}. By contrast, LDS converge as $\log^{d-1}(N) / N$~\cite{bib:bossert}, which results in more rapid decrease according to 
\[
\frac1{N}\,,{\rm ~for~}N\to\infty\,.
\]
This superior convergence rate is achieved by utilizing a sequence of points that {\em per constructionem} sample the unit hypercube as uniformly as possible. Thus, LDS are fully deterministic and not random, despite that they are often referred to as ``quasi-random numbers''. LDS should not be confused with an arrangement of equidistant points on a lattice, which shows a slower convergence rate for typical applications: for example, $N=n^d$ samplings of the integral are needed to fill a lattice with $n$ points per dimension.

The uniformity of the coverage of the unit hypercube can be quantified rigorously by introducing the mathematical concept of discrepancy \dstar. For the purposes of this document, a lower \dstar\ value results in a more uniform coverage of the unit hypercube, and thereby a faster convergence of the MC integration. A rigorous definition and discussion of the \dstar concept is beyond the scope of this document, and can be found in Ref.~\cite{bib:bossert}.

The simplest LDS is given by the van der Corput series in one dimension~\cite{bib:bossert}, which achieves a uniform coverage of the interval $[0,1]$ through a consecutive placement of sampling points at $0,\frac12,\frac14,\frac34,\frac18,\frac58,\frac38,\frac78,$ etc. Among the best performing multidimensional LDS are those given by Faure~\cite{bib:faure}, Sobol~\cite{bib:sobol}, and Niederreiter~\cite{bib:nieder}, which are all based on the van der Corput series. Based on the findings in Ref.~\cite{bib:bossert}, we disregard the Faure sequence. For reducing the computational demand of the ME technique, we tried computer program implementations of the Sobol~\cite{bib:brat88} and Niederreiter~\cite{bib:brat92} sequences provided by the Intel {\sc fortran} compiler~\cite{bib:intel}. Both indicate a similar performance in the convergence rate of the MC integration. Most of our findings presented below apply therefore to both the Sobol and Niederreiter sequences. However, the time for the generation  of the Sobol sequence is considerably less than for the Niederreiter sequence, and we therefore use the Sobol sequence for our implementation of the ME technique, and as the LDS of reference in this document.

One of the central points in numerical integration is to determine reliably the level of achieved precision: an overly optimistic estimate may result in worsened performance of the method because of its greater numerical uncertainty. However, a too pessimistic estimate will waste computing resources. For the numerical evaluation of the integral $G$ of a function $g$ defined on the unit hypercube $[0,1]^d$ using MC integration based on pseudo-random numbers, the standard error estimate is often used:
\begin{equation}
\label{eq:epsstd}
\hat\eps_{\rm std} \equiv \frac1{\sqrt{N}}\left\{ \frac1{N-1}\sum_{i=1}^N\left( g(\xi_i) - \langle g\rangle \right)^2 \right\}^{\frac12}\,,
\end{equation}
where $\xi_i$ within $[0,1]^d$ are the sampling points, and $\langle g\rangle \equiv \frac1{N}\sum_{i=1}^N g(\xi_i)$.
An alternative, rarely used approach, is to split the original sequence of sampling points $\xi_i$ into $K$ sub-sequences with $\frac NK$ sampling points, and make $K$ independent integral estimates $G_k\equiv\frac K{N}\sum_{i=1}^{N/K} g(\xi_{k+(i-1)K})$, $k=1,...,K$. The error estimate is then given by the sample variance of $G_K$, i.e.,
\begin{equation}
\label{eq:eps}
\hat\eps_K \equiv \left\{ \frac1{K-1}\sum_{k=1}^{K}\left( G_k - \langle g\rangle \right)^2 \right\}^\frac12\,,
\end{equation}
where we have chosen $N$ such that $\frac NK$ is an integer. The sum $\sum_{k=1}^K G_k$ follows the Student $t$-statistic~\cite{bib:student}, which approaches the normal distribution in the limit $K\to\infty$. Consequently, for finite $K$, the interval $[\langle g\rangle-\hat\eps_K,\langle g\rangle+\hat\eps_K]$ corresponds to a somewhat smaller confidence level than for the normal distribution.

The error estimator $\hat\eps_{\rm std}$ in Eq.~(\ref{eq:epsstd}) is not appropriate for LDS, as it is too pessimistic. This follows because Eq.~(\ref{eq:epsstd}) applies to \dstar values that are characteristic of pseudo-random numbers, while much smaller \dstar are characteristic of LDS. By constrast, the error estimator $\hat\eps_K$ in Eq.~(\ref{eq:eps}) applies also to LDS, under the condition that each of the $K$ sub-sequences used to obtain independent integral estimates $G_k$ is characterised by the same \dstar\ value as the initial sequence.

However, constructing $K$ independent LDS, with same characteristic \dstar\ values is not trivial: for example, randomly assigning each point of the initial sequence to one of the $K$ sub-sequences results in sub-sequences with characteristic \dstar\ values that are different from that of the initial sequence, and from each other. Several involved and sophisticated approaches have been developed to construct subsequences with the same characteristic \dstar\ value as the initial sequence. For instance, the ME technique implemented as described in Ref.~\cite{bib:cdfme} uses scrambling~\cite{bib:owen98}. For our implementation of the ME technique, we use the ingeniously simple prescription by Warnock~\cite{bib:warnock}. It utilizes the fact that a LDS of points in $[0,1]^{d\cdot K}$ can be regarded as $K$ sub-sequences in $[0,1]^d$, which have the same \dstar value {\em per constructionem}. 
For our implementation, with $d=10$, we generate one LDS of $d=40$, i.e., $K=4$, which offers a reliable error estimate at a confidence level of about 63\%.

Before implementing the LDS in the ME technique, we evaluate their performance and the applicability of error estimators using toy MC integrations of multidimensional test functions: the normal distribution in up to ten dimensions,
and a trigonometric  function inside a torus of three dimensions. In particular, we define the integrand as $g(\rho)\equiv1+\cos(\pi\rho^2/R^2)$, for $\rho\leq r$, and otherwise $g(\rho)=0$, where $\rho$ is the distance from a given sampling point to the center of the torus tube, with $R=0.6$ being the distance from the center of the tube to the center of the torus, and $r=0.3$ the radius of the tube. The integration volume is $V\equiv[-1,1]^3$. In the following, we focus on the numerically more challenging example of the trigonometric function inside a three-dimensional torus.
As a figure of merit, we use 
\begin{equation}
{\rm relative~convergence}\equiv\frac{|\langle g\rangle\cdot V-\int_Vf\dif V|}{\int_Vf\dif V}\,,
\label{eq:conv}
\end{equation}
where $V$ is the integration volume, and $\int_Vf\dif V=2\pi^2Rr^2$ is the analytic result.

\begin{figure}
\begin{centering}
\includegraphics[width=0.99\columnwidth]{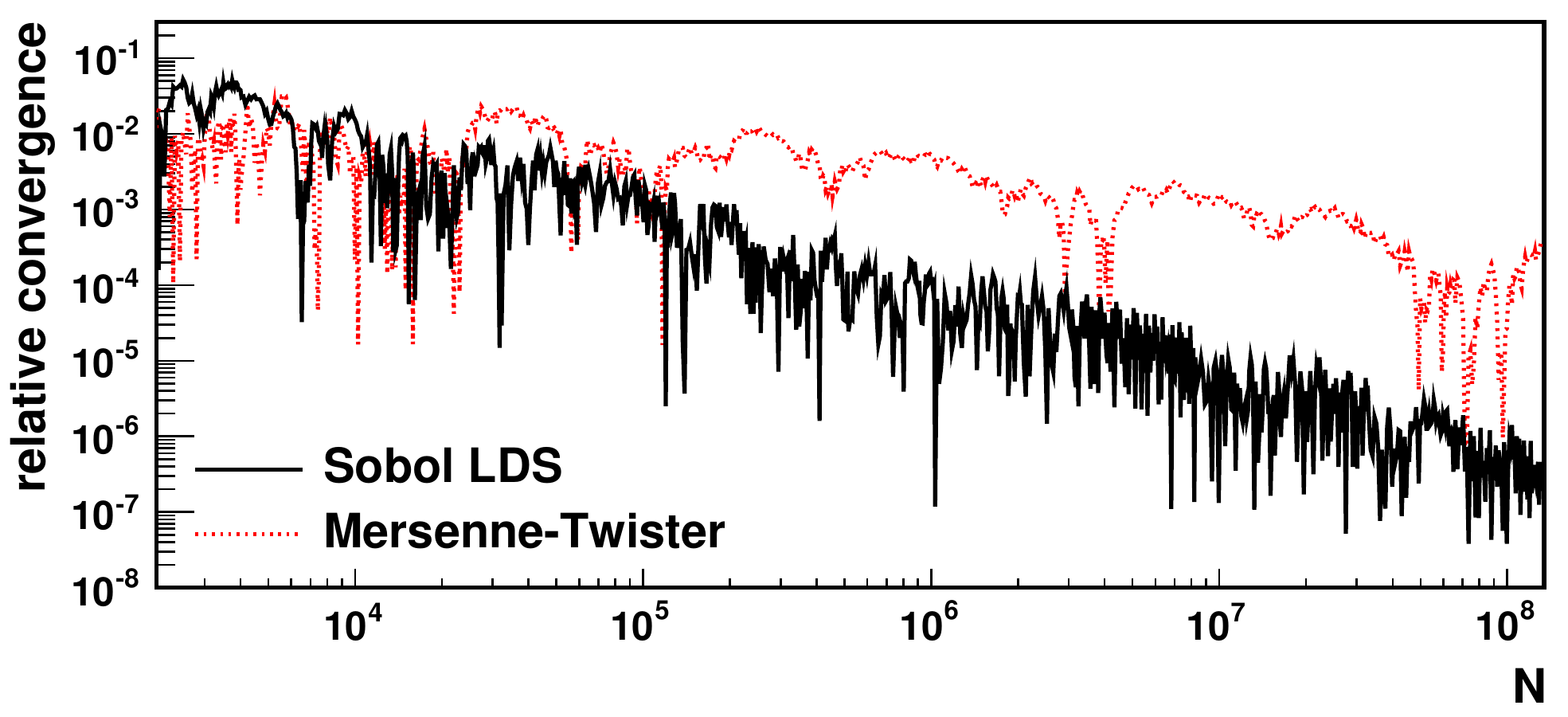}
\par\end{centering}
\caption{
\label{fig:conv}
The relative convergence for numerical evaluation of the integral of a trigonometric function inside a torus in three dimensions. The results are obtained using the MC integration technique based on the Sobol LDS and the Mersenne-Twister pseudo-random number sequence.
}
\end{figure}

The relative convergence is compared for the Sobol LDS and the Mersenne-Twister~\cite{bib:mt} pseudo-random number sequence in Fig.~\ref{fig:conv} for $N$ up to $1.3\times10^{8}$. As anticipated, the Sobol sequence displays superior convergence behaviour that follows $\frac1N$. Taking $N=2^{26}\approx6.7\times10^{7}$ as an example, the Sobol sequence outperforms the Mersenne-Twister sequence by more than 3 orders of magnitude, and achieves a relative convergence of $5.7\times10^{-7}$ compared to $1.0\times10^{-3}$.

\begin{figure}
\begin{centering}
\includegraphics[width=0.99\columnwidth]{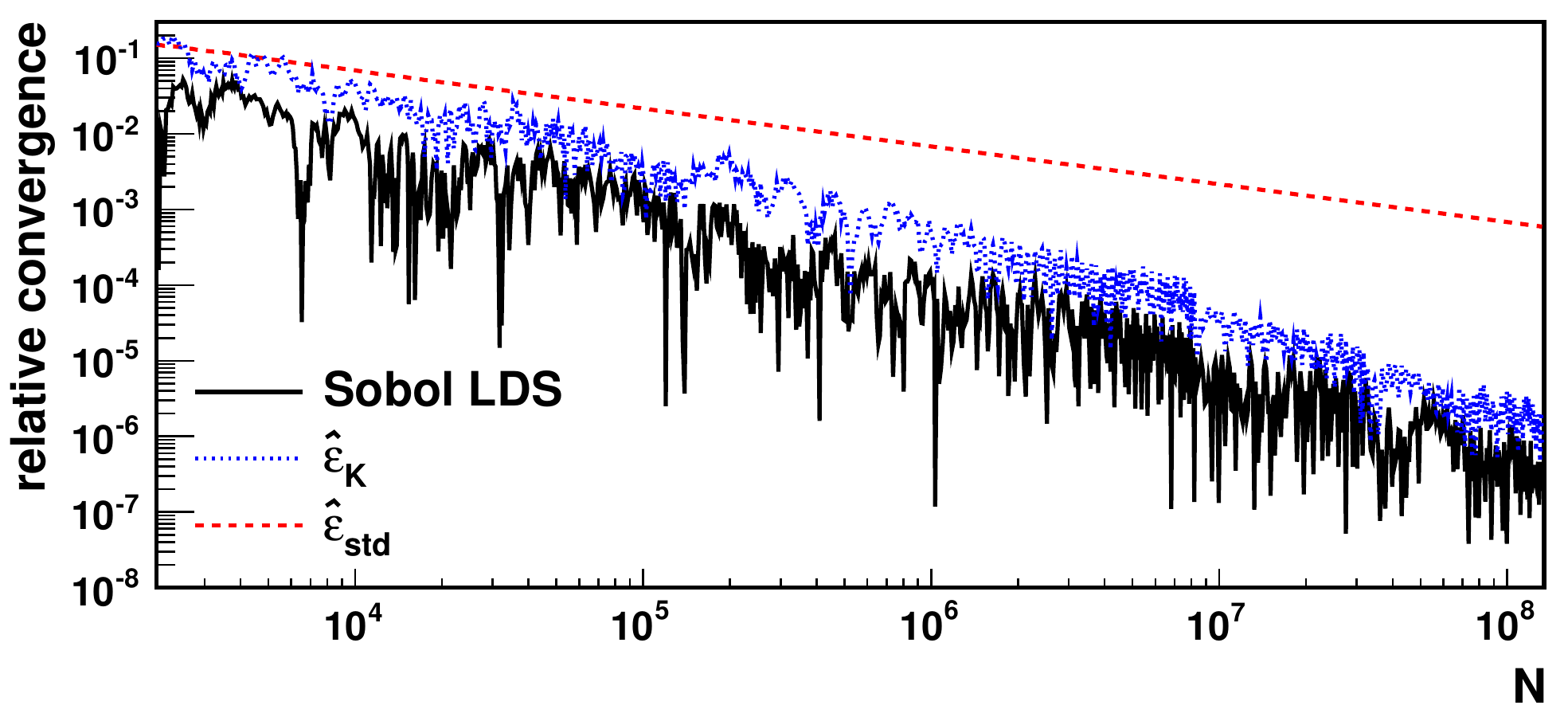}
\par\end{centering}
\caption{
\label{fig:err}
The relative convergence for numerical evaluation of the integral of a trigonometric  function inside a torus in three dimensions.  The results are obtained using the MC integration technique based on the Sobol LDS. Also shown are the error estimates $\hat\eps_K$ from Eq.~(\ref{eq:eps}) for $K=4$ and $\hat\eps_{\rm std}$ from Eq.~(\ref{eq:epsstd}).
}
\end{figure}

\begin{figure}
\begin{centering}
\includegraphics[width=0.99\columnwidth]{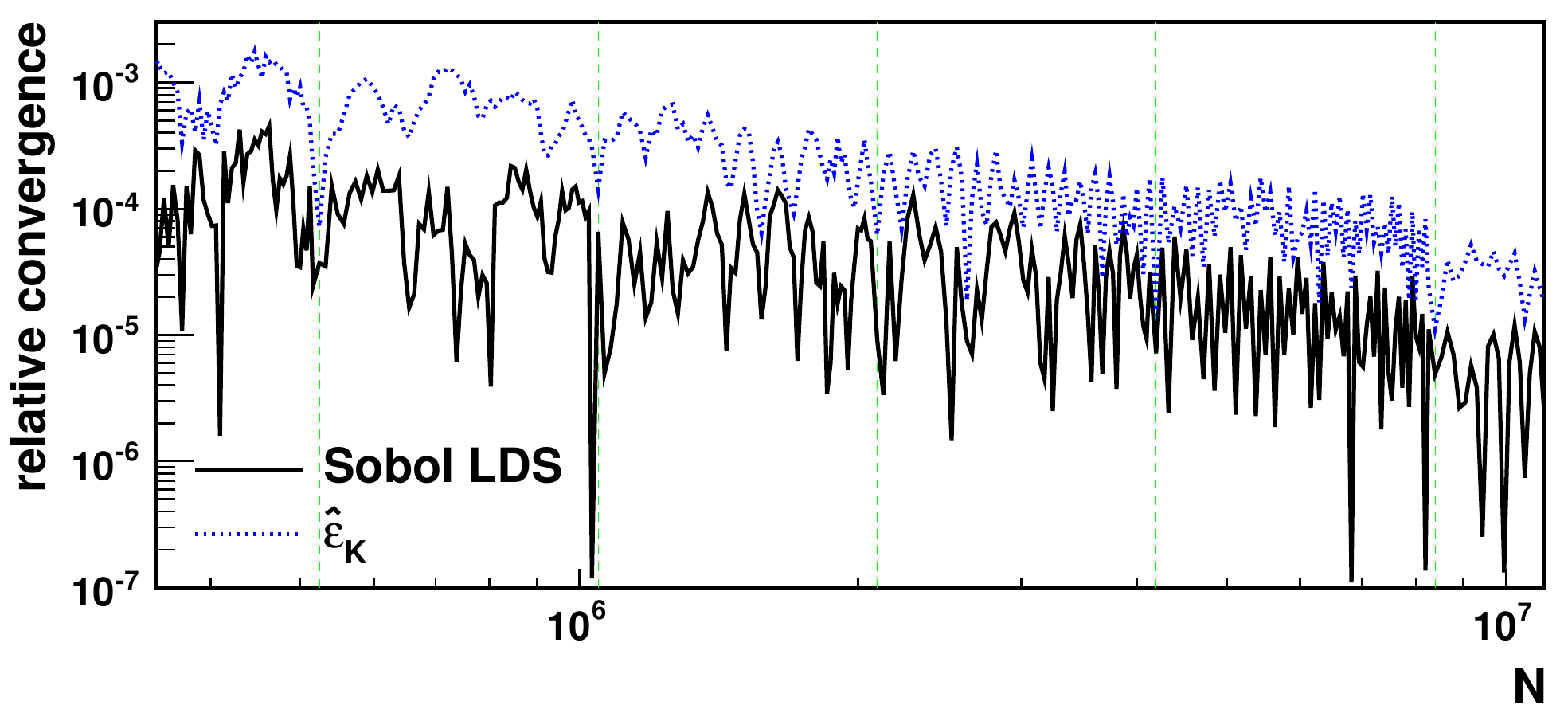}
\par\end{centering}
\caption{
\label{fig:errzoom}
Same as Fig.~\ref{fig:err}, however for a sub-range of $N$, and without showing $\hat\eps_{\rm std}$. Powers of 2 are indicated by the thin vertical broken lines in green.
}
\end{figure}

The performance of the error estimates  $\hat\eps_K$ from Eq.~(\ref{eq:eps}) for $K=4$ and $\hat\eps_{\rm std}$ from Eq.~(\ref{eq:epsstd}) using the trigonometric function inside a three-dimensional torus as a test function and the Sobol LDS is evaluated in Fig.~\ref{fig:err}. Evidently, $\hat\eps_{\rm std}$ yields a too pessimistic error estimate for the Sobol sequence, despite that it is applicable to the Mersenne-Twister sequence, as can be seen from comparison with Fig.~\ref{fig:conv}. 
By contrast, $\hat\eps_K$ provides an appropriate error estimate for the Sobol sequence, and can therefore be used in our implementation of the ME technique. A practical feature of the $\hat\eps_K$ estimator is that it dynamically follows the relative convergence, i.e., $\hat\eps_K$ tends to be small for small values of the relative convergence. This is not the case for $\hat\eps_{\rm std}$, which merely gives a monotonously falling upper bound. Thus, $\hat\eps_K$ can provide a dynamic indication of the achieved numerical precision through the dips observed in the relative convergence. This feature of $\hat\eps_K$ is illustrated for $K=4$ in Fig.~\ref{fig:errzoom}, for a subrange in $N$. We remark that the dips in relative convergence and, consequently, $\hat\eps_K$ tend to occur for $N=2^n$, where $n$ is an integer. This is because the unit hypercube is sampled most uniformly for such $N$. We profit from this feature in various places of our implementation of the ME technique, for example, when we perform the pre-integration step (described in Sec.~\ref{sec:previous}) with $N=2^{14}$ samplings of the integral.

Having tested the performance of LDS and verified the applicability of the error estimator $\hat\eps_{K=4}$, we proceed to implement the Sobol sequence in our ME technique. As described in Sec.~\ref{sec:previous}, we use prior information for most of the integration variables by using sampling points distributed according to the prior function, a technique that is commonly referred to as importance sampling. Two common approaches to achieve this are the accept-reject method, and the cumulative distribution function (CDF) method based on the CDF of the prior~\cite{bib:numrec}. It is important to note that the former method, previously employed in Ref.~\cite{bib:mtop36}, cannot be used in conjunction with LDS. This is because LDS loses its superior property of a low \dstar value if a subset of points of the sequence is rejected. By contrast, in the CDF method, all points are preserved and only their mapping from the unit hypercube to the sampling space is modified. No other major changes are required for using the Sobol sequence in our implementation of the ME technique besides switching to the CDF method for sampling the integration space according to a given prior.

After implementing the Sobol sequence for MC integration, we find a reduction of the computation time for the calculation of $\psig$ from about 2~hours per event, averaged over the sample of simulated \ttbar events for $\mt=172.5~\GeV$, to about 15~min/event, i.e., by about one order of magnitude. This improvement is for a required numerical precision of 1\%, which is found to be sufficient for a robust statistical performance of our implementation. However, our tests with MC integration indicate that the relative gain in computation time may be even greater for smaller required precision. 

\section{
Reducing the computation demand by factoring out the $\kjes$ dependence from the matrix element calculation
\label{sec:kjes}
}

As already mentioned in Sec.~\ref{sec:previous}, we construct the likelihood on a grid in $(\mt,\kjes)$ with spacings of $(1~\GeV,0.01)$ for the extraction of \mt and \kjes. For standard samples of simulated MC events which account for a major fraction of the computational demand, this is done for \mt within $[\mtgen-12~\GeV,\mtgen+12~\GeV]$ and for \kjes within $[\kjes^{\rm gen}-0.1,\kjes^{\rm gen}+0.1]$, where $\mtgen$ is the generated $\mt$ and $\kjes^{\rm gen}$ the generated $\kjes$. 
Thus, \psig has to be calculated for $25\times21=525$ grid points in $(\mt,\kjes)$.

In our previous implementation of the ME technique, we recalculated \psig entirely for each point in $(\mt,\kjes)$. 
However, the integrand in Eq.~(\ref{eq:psig}) depends on \kjes only via the transfer function $W(\x, \y,\kjes)$. Furthermore, as detailed in Sec.~\ref{sec:previous}, the integration in Eq.~(\ref{eq:psig}) is performed over nine partonic variables and $\kappa$, none of which depend on \kjes. Therefore, in our new implementation, we factor out the $\kjes$ dependence from the ME computation and perform the calculation of $\mu(\y;\mt)\equiv\sum \dif\sigma(\y;\mt)\dif \vec q_1\dif\vec q_2 f(\vec q_1)f(\vec q_2)$ in Eq.~(\ref{eq:psig}) only once for a given sampling point. We then obtain all the 21 integrand values in Eq.~(\ref{eq:psig}) for the different $\kjes^i$, $i=1,2,...,21$ by multiplying $\mu(\y;\mt)$ with the transfer function $W(\x, \y,\kjes^i)$. Thus, we obtain 21 simultaneous estimates for $\psig$.

After factoring out the \kjes dependence from the ME calculation as described above, we find a further reduction of the computation time for the calculation of \psig from about 15~min/event, after the implementation of LDS, to about 80~s/event, i.e., by another order of magnitude. We note that the reduction is somewhat smaller than the factor of 21 that would be naively expected from the number of grid points in \kjes. This is because of the increased overhead of keeping track of the 21 simultaneous \psig estimates.

The computation time for \pbkg is much less of an issue compared to \psig. This is because \pbkg does not depend on \mt by definition, and has to be calculated only for 21 points in \kjes. Therefore, we did not apply the new approach of factoring out the \kjes dependence from the matrix element calculation in \pbkg.


\section{
Validation of the new implementation of the matrix element technique
\label{sec:val}
}
To verify that the sensitivity of our implementation of the ME technique was not adversely affected by the modifications described in Secs.~\ref{sec:lds} and~\ref{sec:kjes}, we study the response of the ME technique in \mt and \kjes. This is done by comparing the extracted $\mtfit$ with the generated $\mtgen$ using pseudo-experiments, and using analogous procedures for $\kjes$. The pseudo-experiments are comprised of \ttbar events and dominant background contributions according to their respective fractions measured in data~\cite{bib:mtop}. To evaluate the method's response in \mt, we use five simulated MC samples for \ttbar production with $\mtgen=165,170,172.5,175,$ and $180~\GeV$ for $\kjes^{\rm gen}=1$. Similarly, for \kjes we use signal and background MC samples with $\kjes^{\rm gen}=0.95,1,1.05$, and \ttbar signal is generated for $\mtgen=172.5$. In this validation, we study a representative set of simulated samples used to model data corresponding to 3~\fb of integrated luminosity.


\begin{figure}
\begin{centering}
\includegraphics[width=0.49\columnwidth]{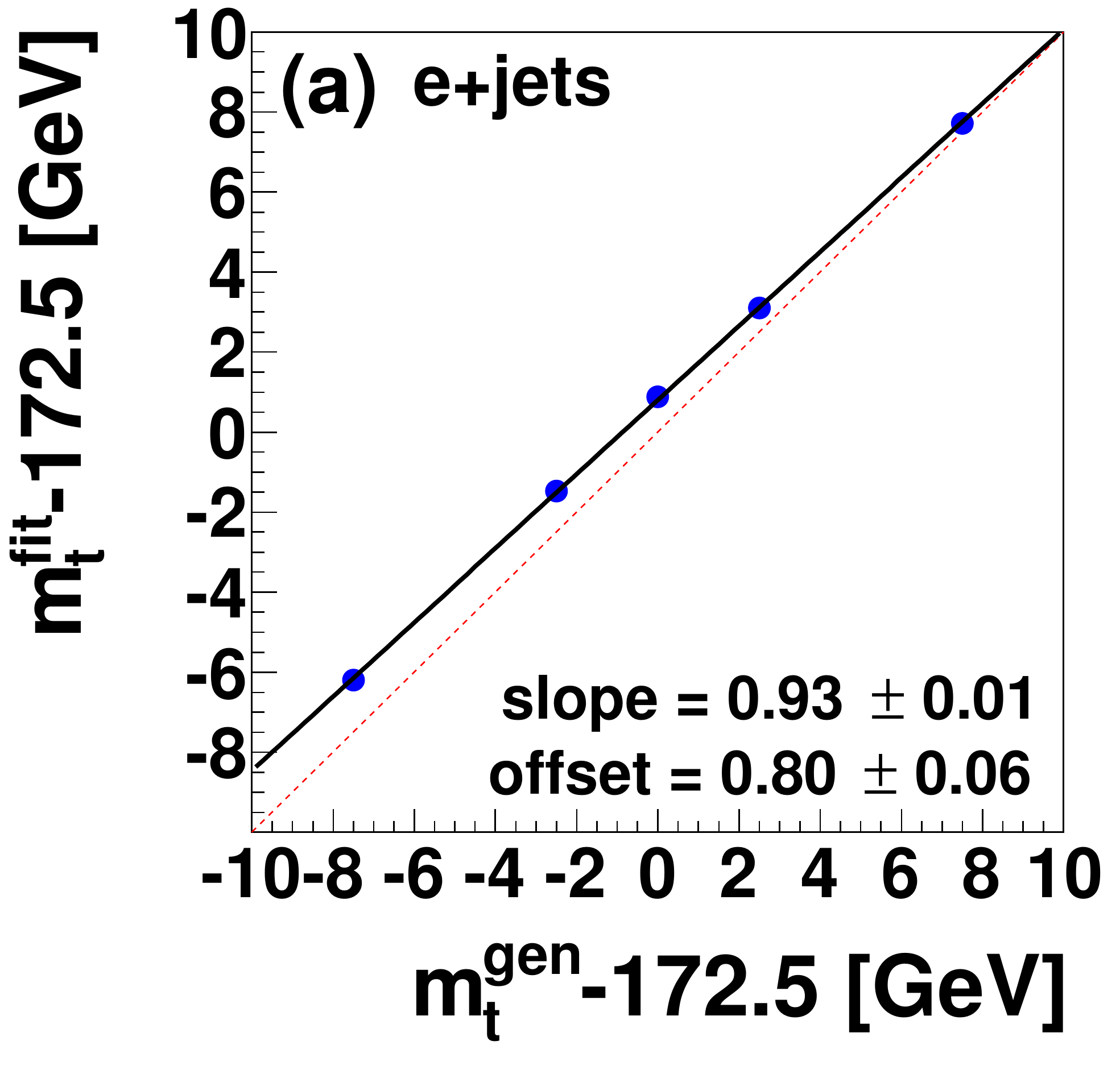}
\includegraphics[width=0.49\columnwidth]{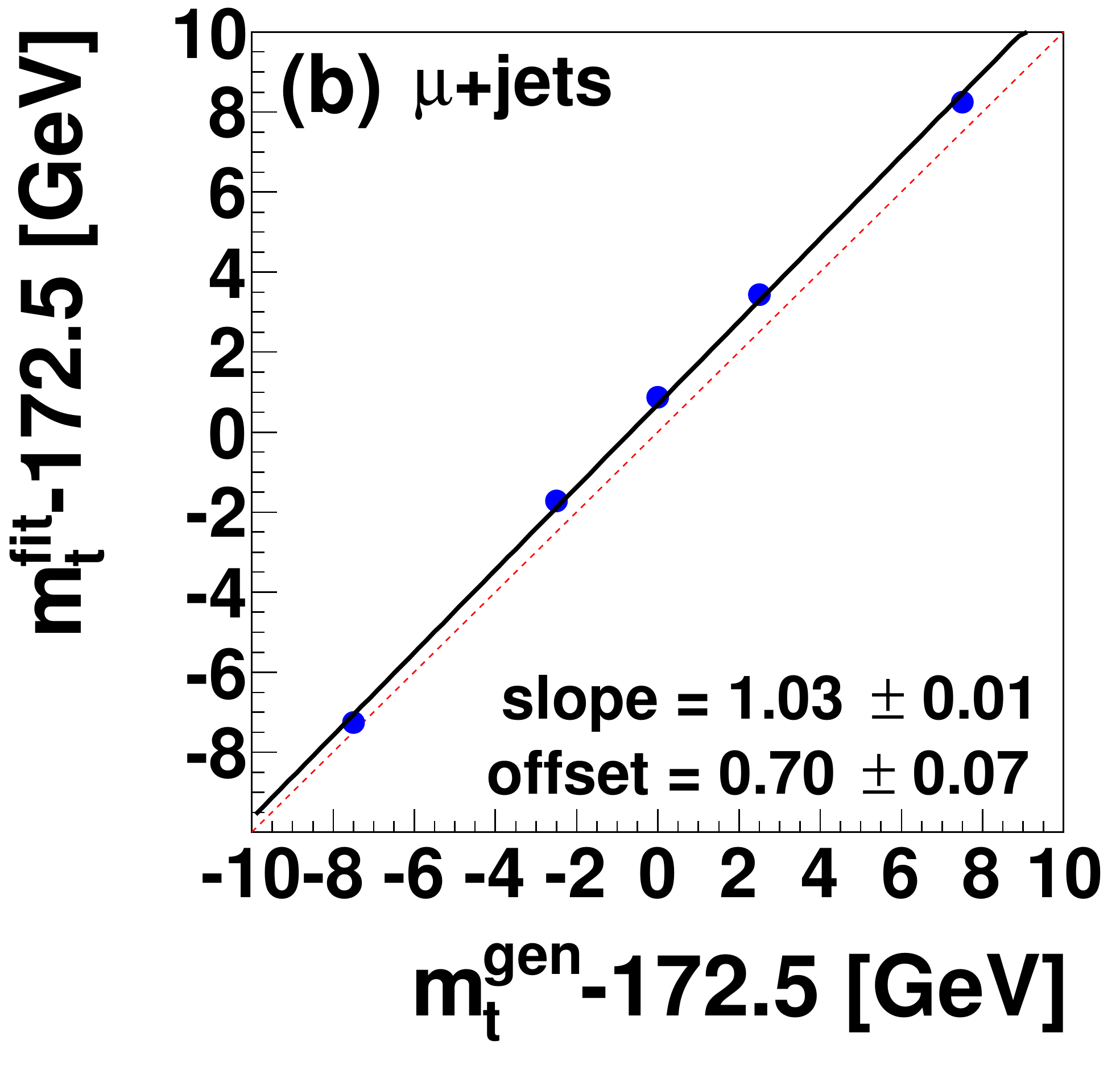}
\includegraphics[width=0.49\columnwidth]{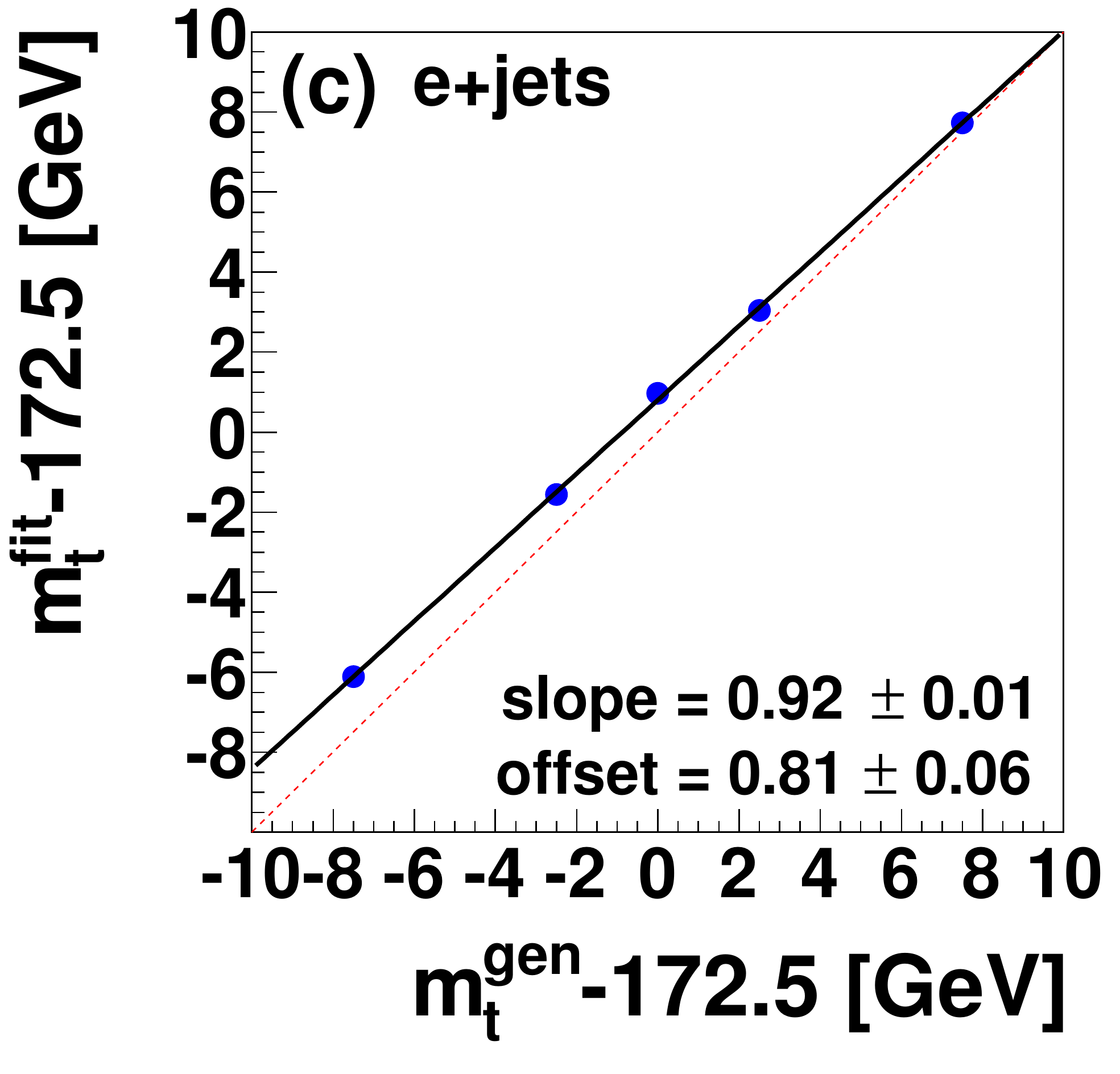}
\includegraphics[width=0.49\columnwidth]{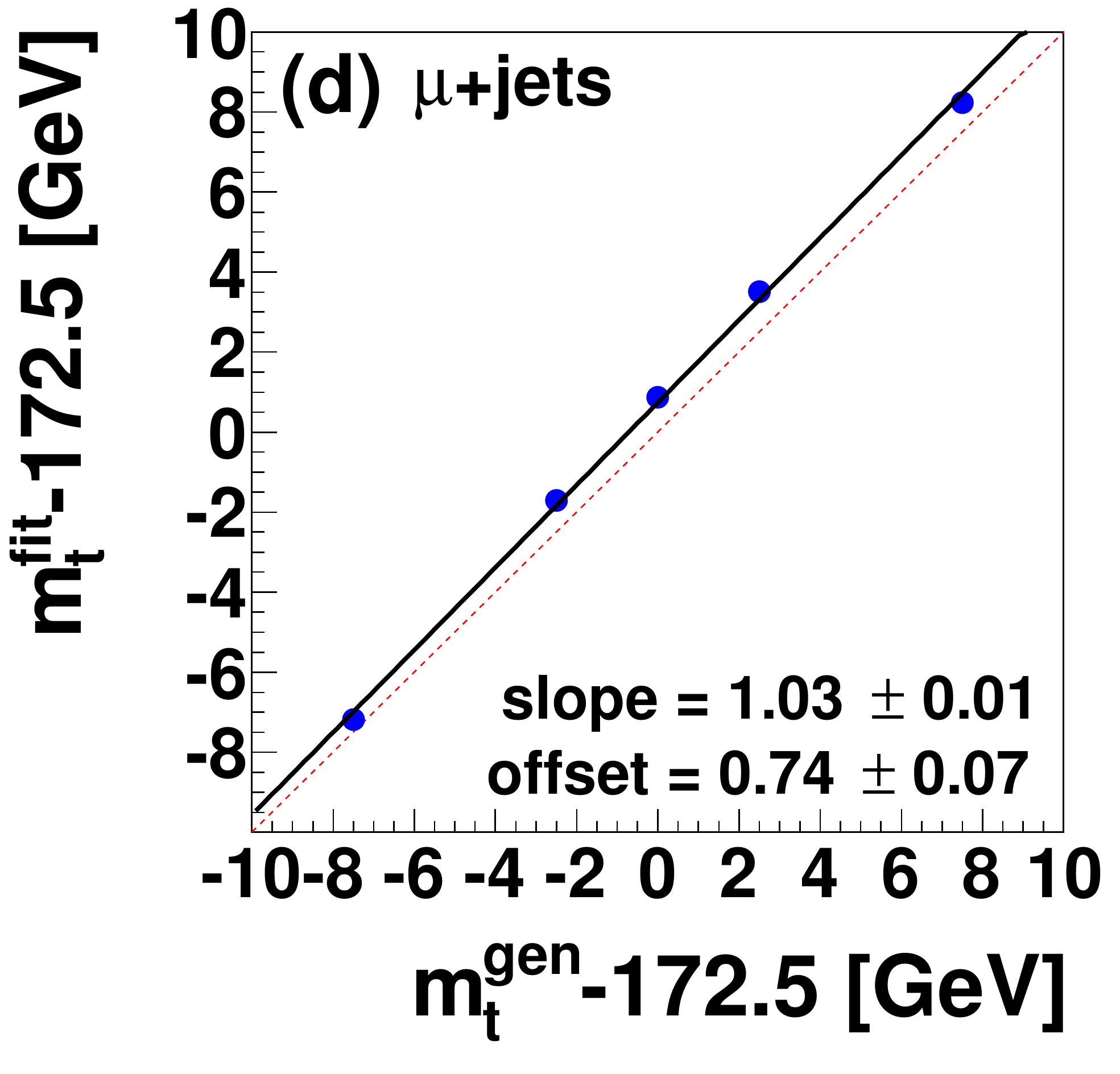}
\par\end{centering}
\caption{
\label{fig:responsemt}
The response of the ME technique in \mt obtained using pseudo-experiments constructed from MC events with fully simulated response of the D0 detector. Each data point corresponds to the mean extracted \mt averaged over 1000 pseudo-experiments at a given \mtgen. The dependence is fitted with a linear function (black solid line), with the ideal case indicated as the red broken line. The results obtained with our previous implementation of the ME technique are shown in (a) for the \ejets and in (b) for the \mujets channel. Analogous results obtained including the improvements described in this manuscript are shown in~(c) and~(d).
}
\end{figure}

\begin{figure}
\begin{centering}
\includegraphics[width=0.49\columnwidth]{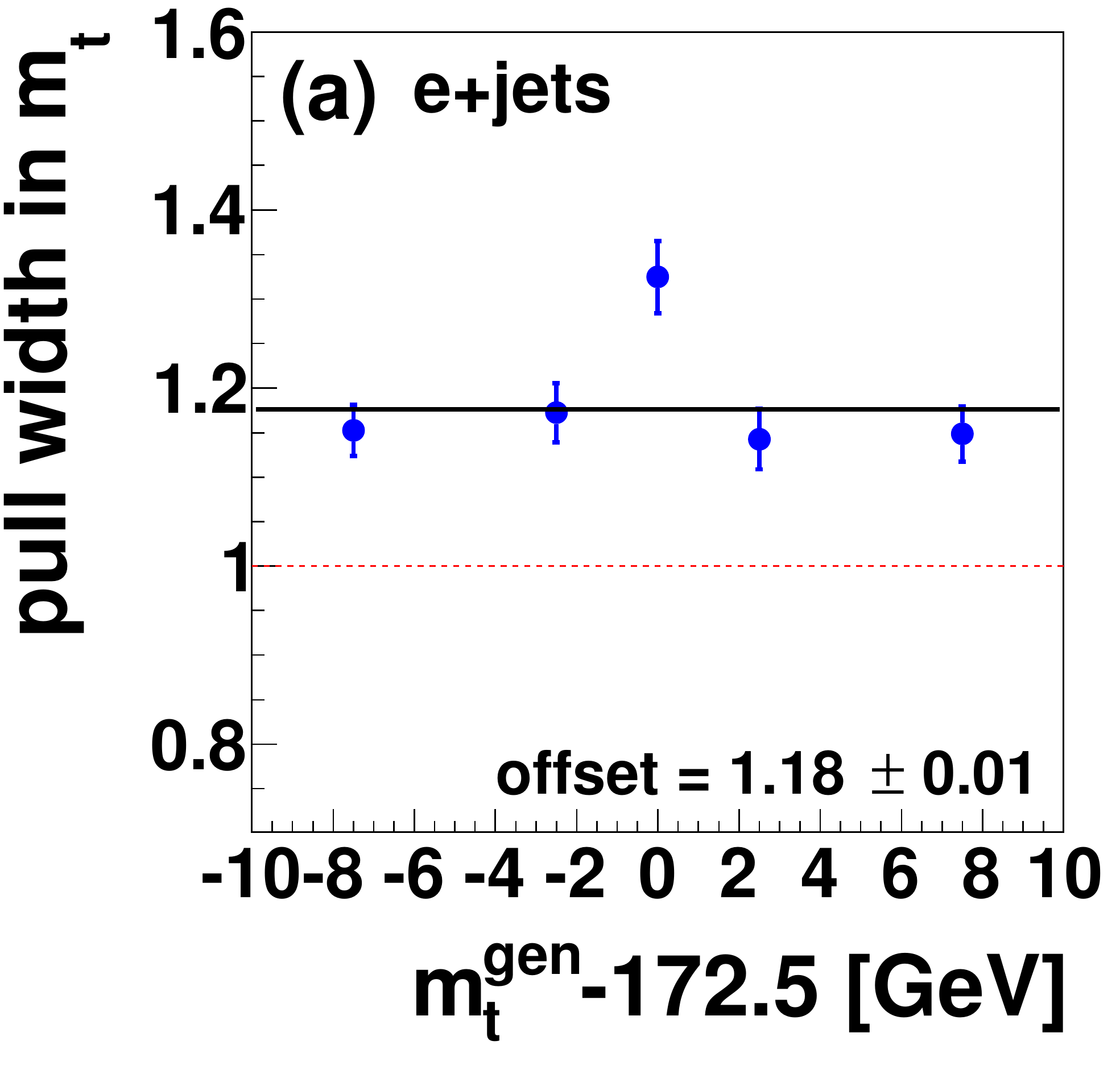}
\includegraphics[width=0.49\columnwidth]{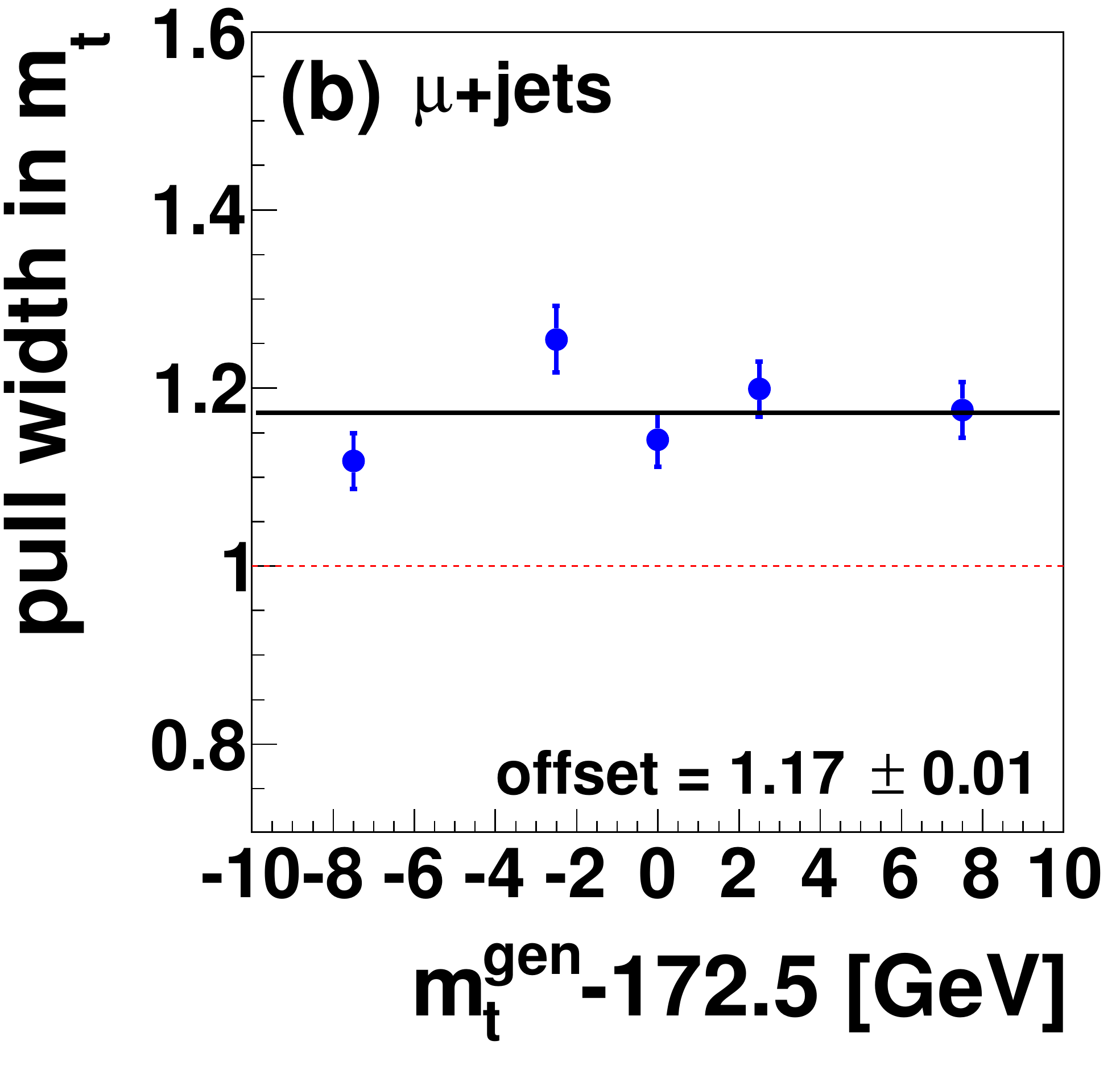}
\includegraphics[width=0.49\columnwidth]{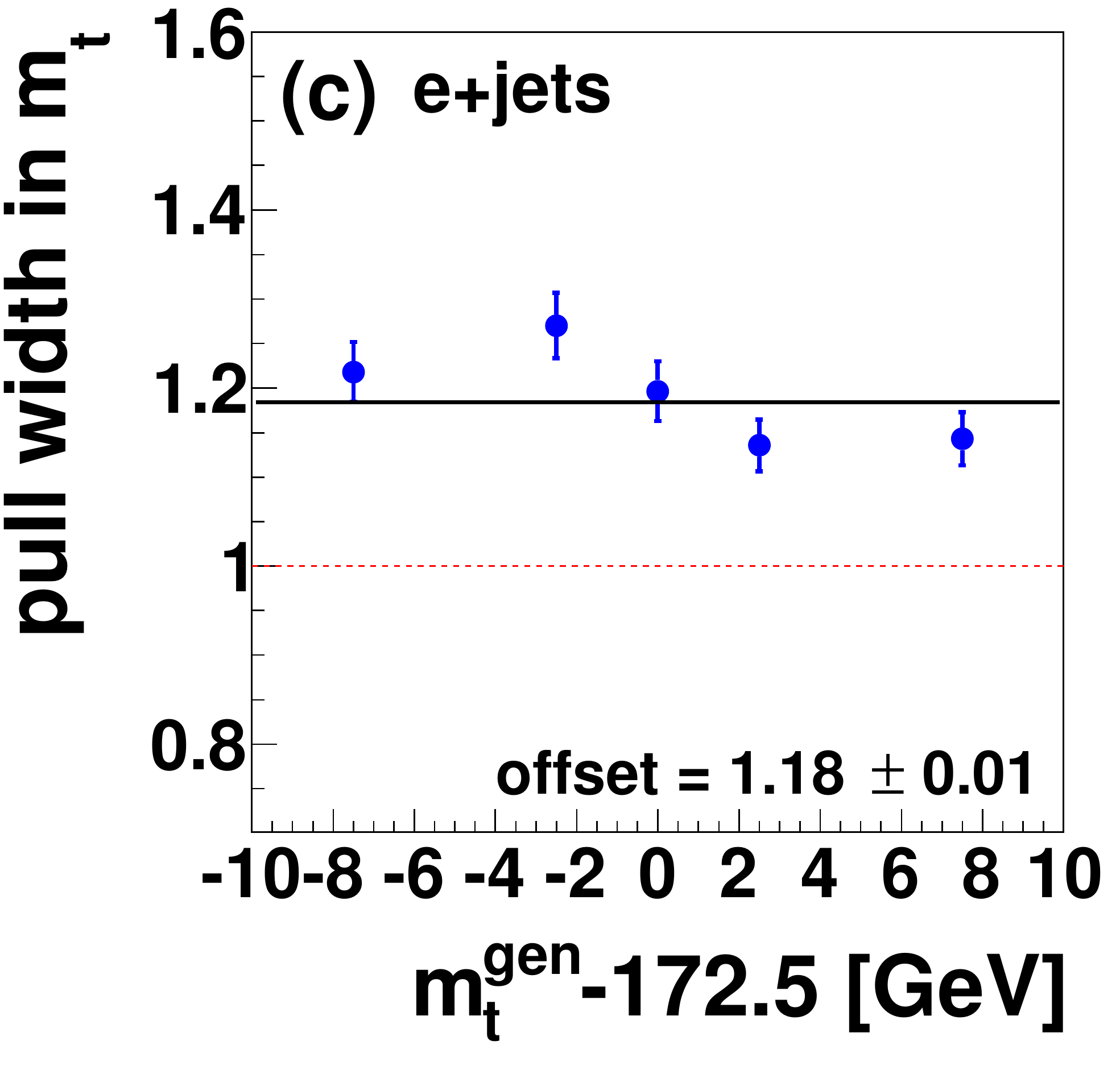}
\includegraphics[width=0.49\columnwidth]{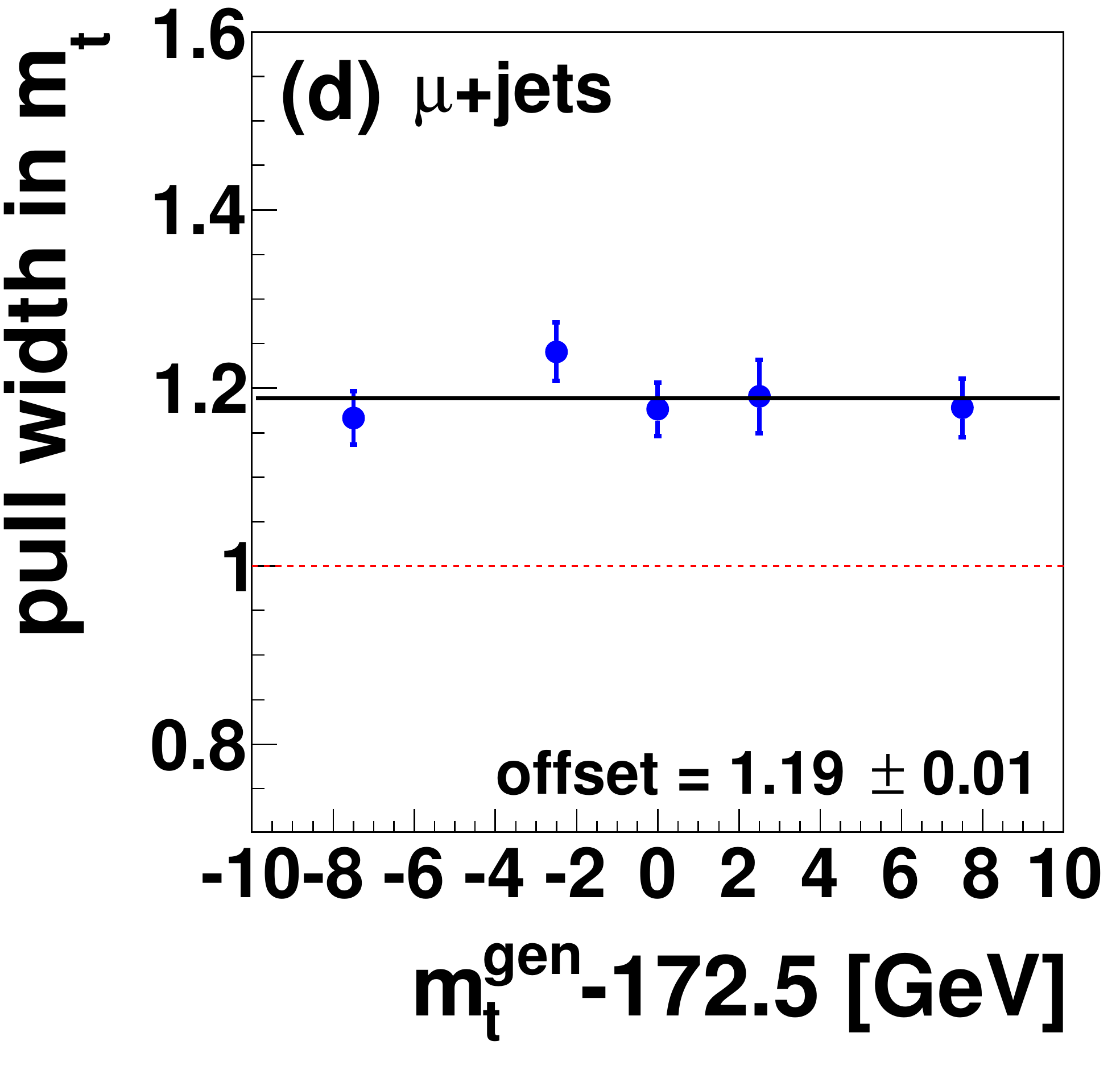}
\par\end{centering}
\caption{
\label{fig:pullmt}
The statistical sensitivity of the ME  technique in \mt obtained using pseudo-experiments constructed from MC events with fully simulated response of the D0 detector. Each data point corresponds to the width of the distribution in the pull of \mt found with 1000 pseudo-experiments at a given \mtgen. The dependence is fitted with a constant (black solid line), with the ideal case indicated as the red broken line. The results obtained with our previous implementation of the ME technique are shown in (a) for the \ejets and in (b) for the \mujets channel. Analogous results obtained including the improvements described in this manuscript are shown in~(c) and~(d).
}
\end{figure}

The response of our implementation of the ME technique in \mt is presented in Fig.~\ref{fig:responsemt}, before and after the improvements described in Secs.~\ref{sec:lds} and~\ref{sec:kjes}. The results are shown split into \ejets and \mujets channels defined by the presence of one isolated electron or muon with $\pt>20~\GeV$. Given that the ME technique calculates event probabilities {\em ab initio} and relies on an analytic parametrization of detector response, its performance is remarkably close to ideal, defined by an offset parameter of zero and by a slope parameter of unity. Most important, the response of the ME technique before and after the improvements 
is consistent within statistical uncertainties.

As an additional cross-check, we verify that the statistical sensitivity of the ME technique remains consistent. For this, we study the width of the distribution in the pull of \mt. The pull of \mt is defined as $\sum_{i=1}^{1000}(\mt^i-\langle\mt\rangle)/\sigma_{\mt}^i$, where $\mt^i$ and $\sigma_{\mt}^i$ are the extracted values of \mt and its corresponding statistical uncertainty found in pseudo-experiment $i$, $\langle\mt\rangle\equiv\frac1{1000}\sum_{i=1}^{1000}\mt^i$, and the sums extend over all 1000 pseudo-experiments conducted for a given \mtgen. As Fig.~\ref{fig:pullmt} demonstrates, the statistical sensitivity is within 20\% of the ideal pull width of unity, and is consistent within statistical uncertainties before and after the improvements.

\begin{figure}
\begin{centering}
\includegraphics[width=0.49\columnwidth]{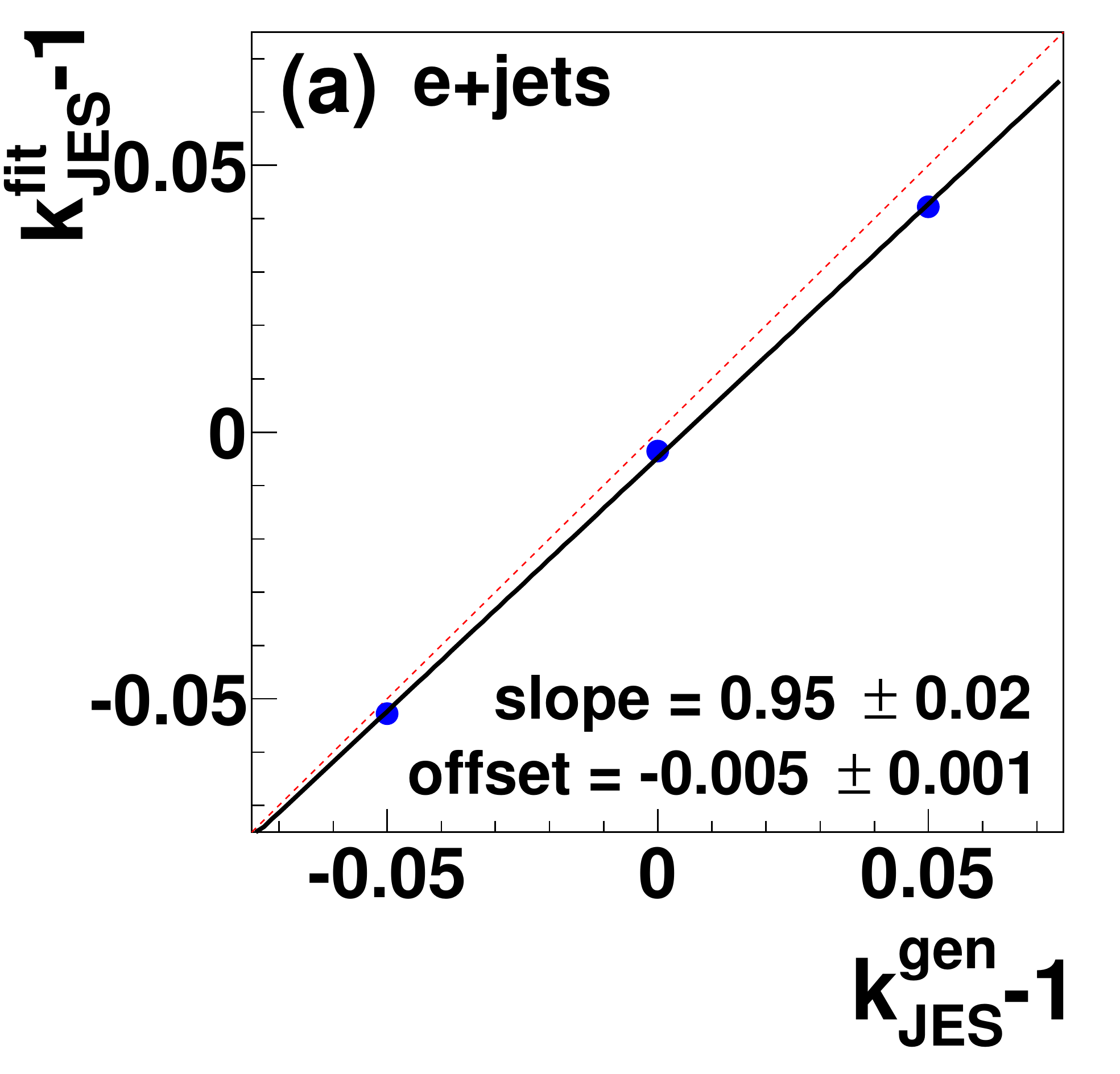}
\includegraphics[width=0.49\columnwidth]{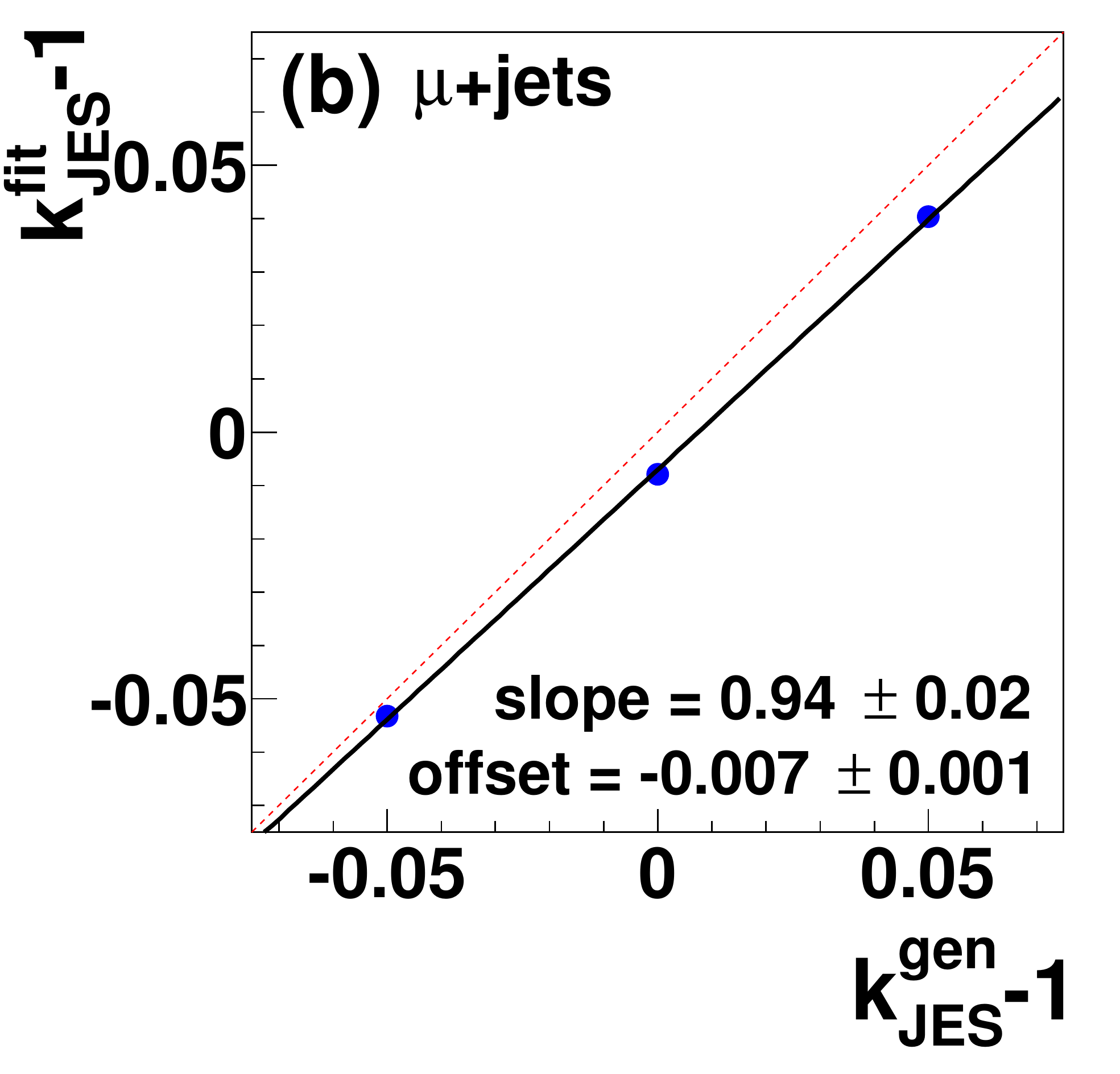}
\includegraphics[width=0.49\columnwidth]{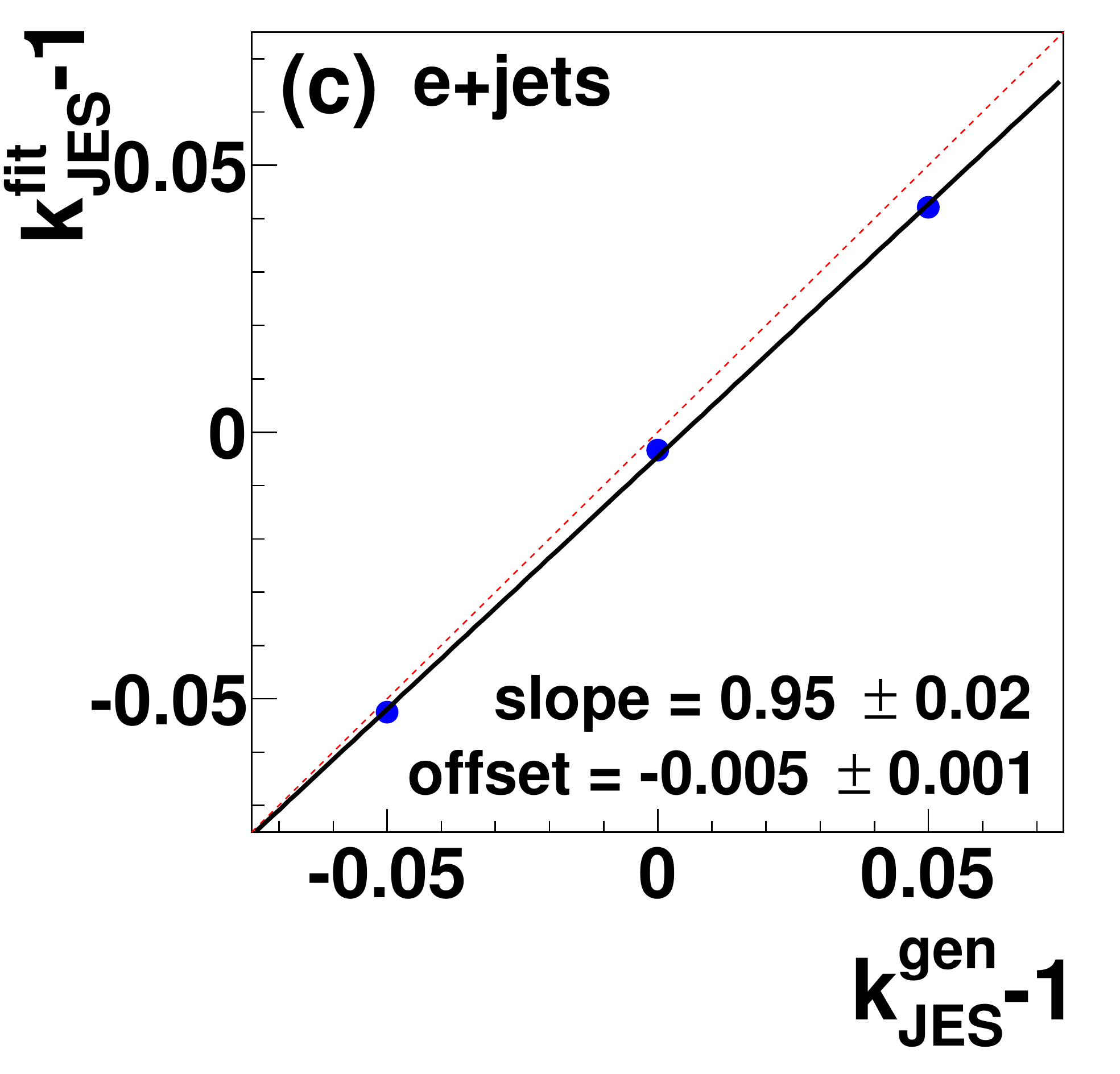}
\includegraphics[width=0.49\columnwidth]{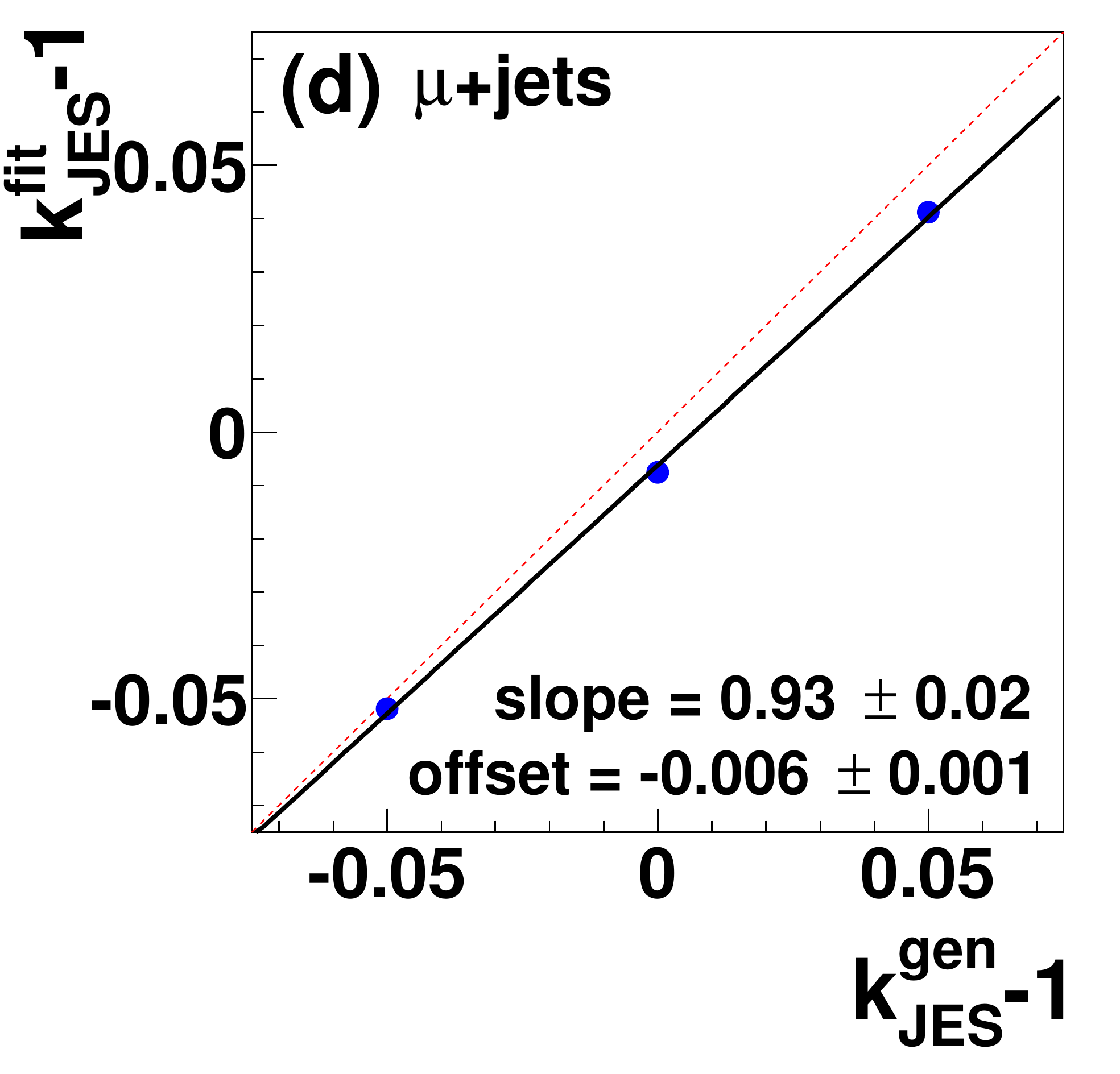}
\par\end{centering}
\caption{
\label{fig:responsekjes}
Same as Fig.~\ref{fig:responsemt}, but for \kjes.
}
\end{figure}

\begin{figure}
\begin{centering}
\includegraphics[width=0.49\columnwidth]{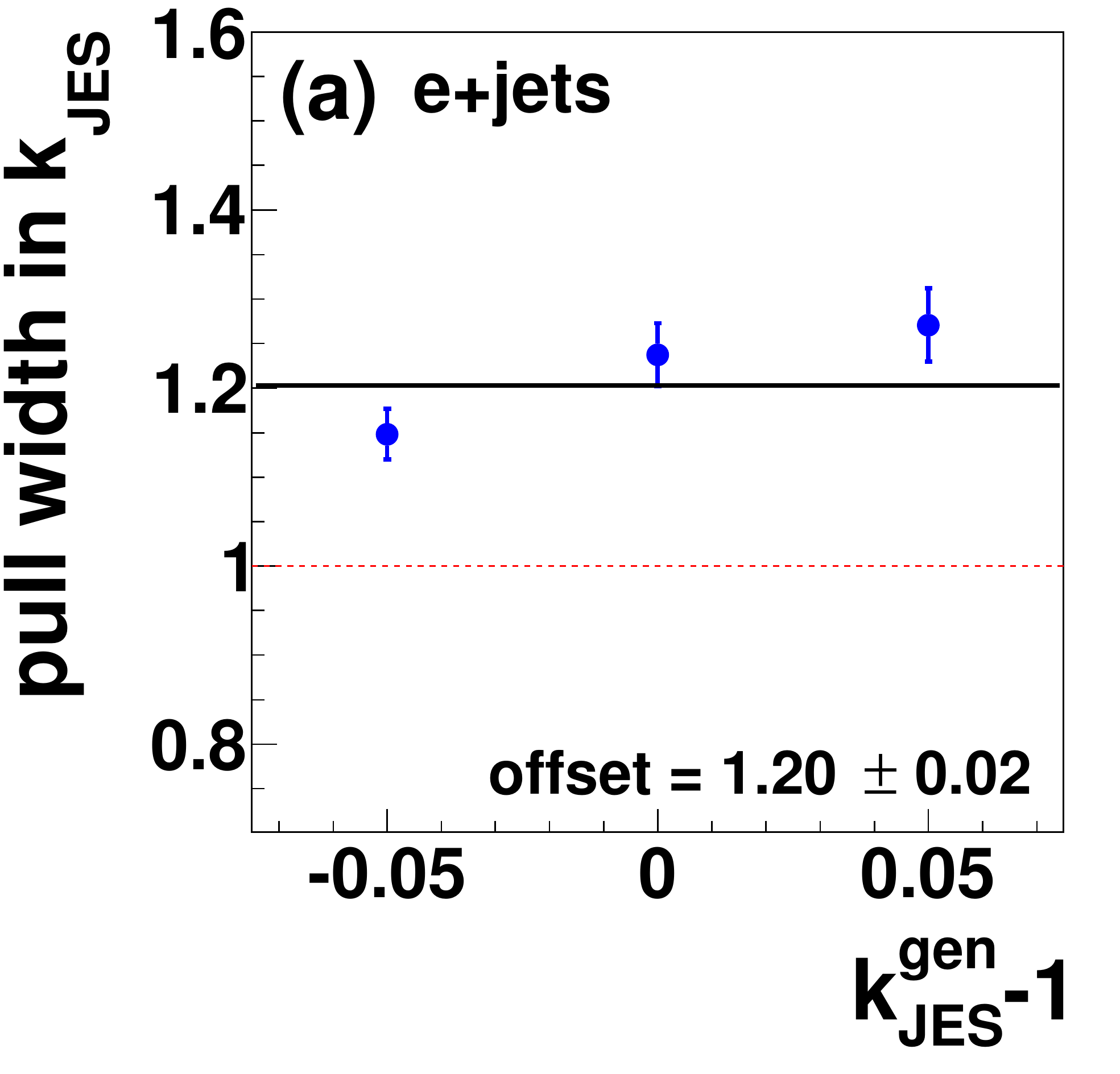}
\includegraphics[width=0.49\columnwidth]{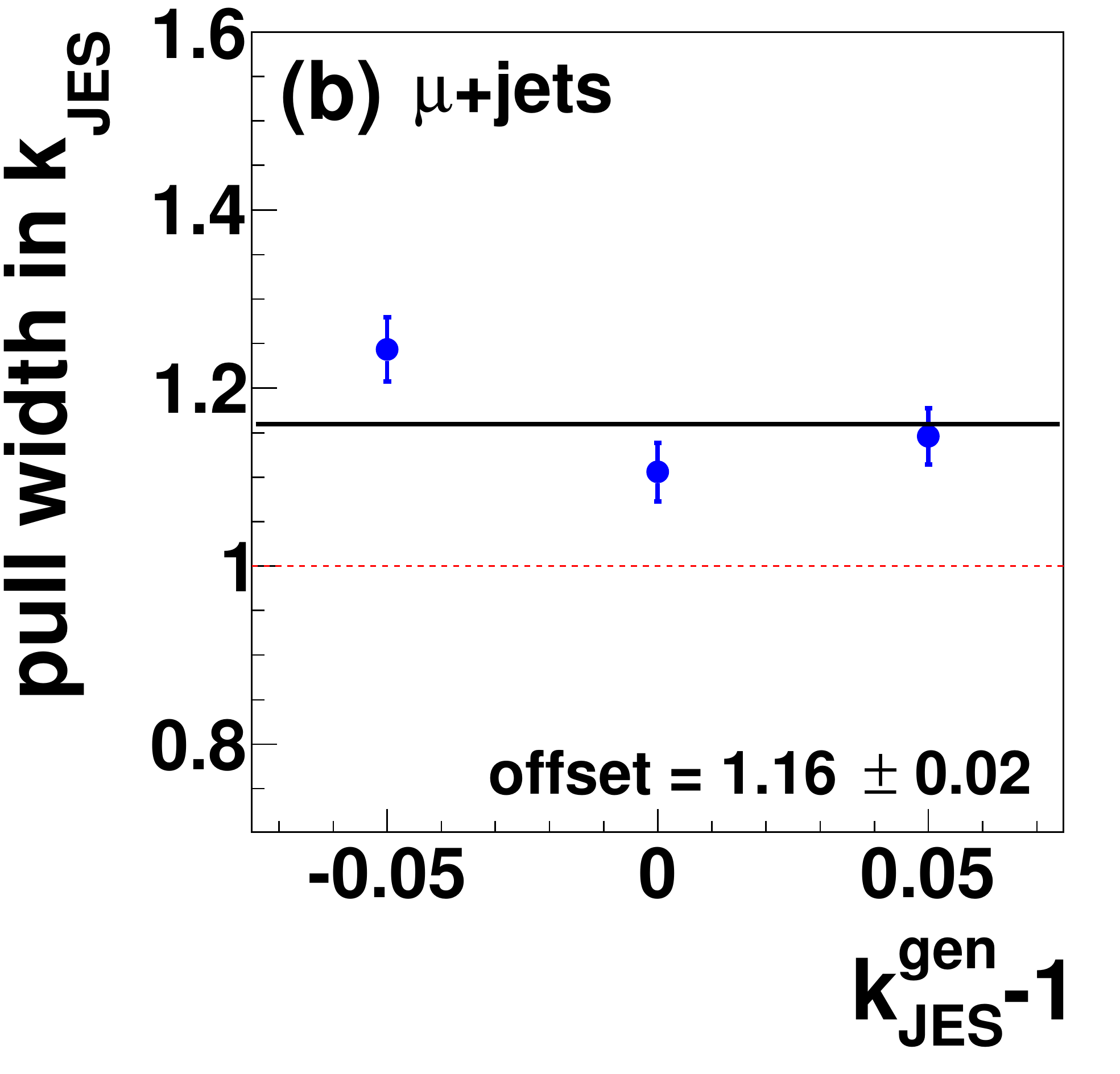}
\includegraphics[width=0.49\columnwidth]{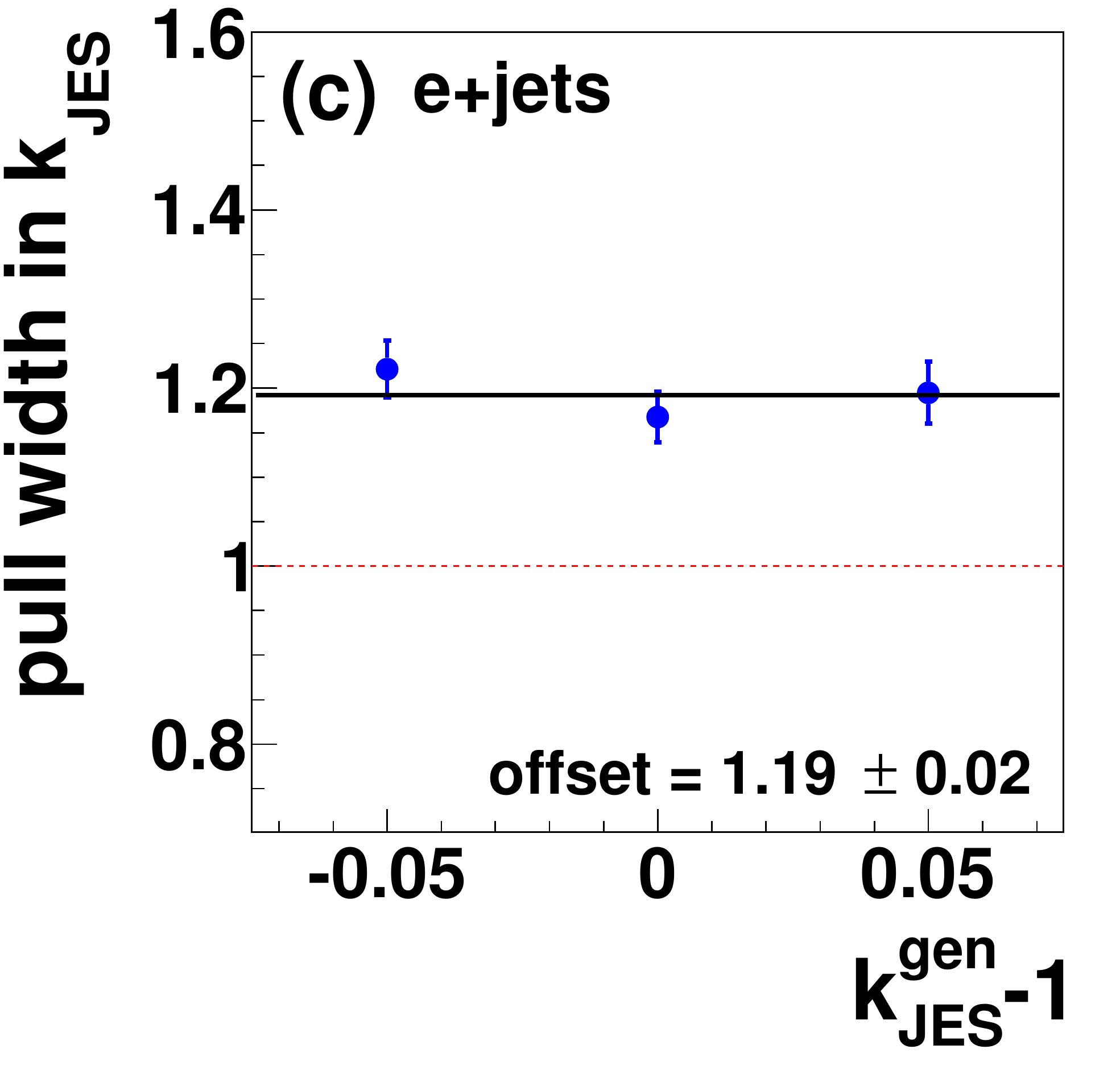}
\includegraphics[width=0.49\columnwidth]{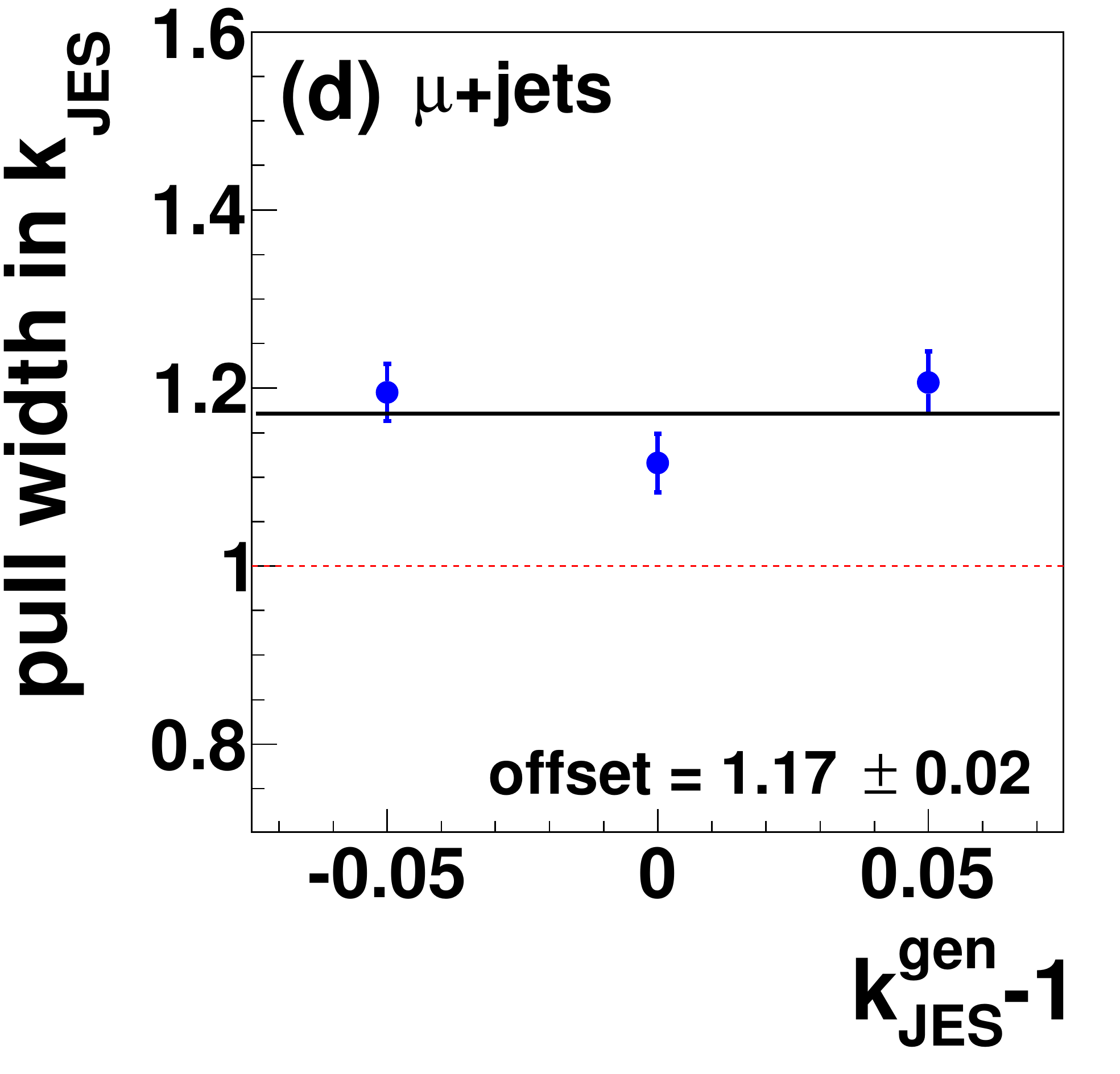}
\par\end{centering}
\caption{
\label{fig:pullkjes}
Same as Fig.~\ref{fig:pullmt}, but for \kjes.
}
\end{figure}

In similar spirit, we study the response and statistical sensitivity in \kjes 
before and after the improvements of Secs.~\ref{sec:lds} and~\ref{sec:kjes}. The results for the response are shown in Fig.~\ref{fig:responsekjes}, while the statistical sensitivity is presented in Fig.~\ref{fig:pullkjes}. Both figures display a consistent performance of the ME technique before and after our improvements.

Our validation studies in \mt and \kjes
indicate full consistency within statistical uncertainties. We therefore conclude that our implementation of the ME technique using LDS for the MC integration and factorizing the \kjes dependence from the ME calculation does not adversely affect the performance of the method, and can be applied for data analysis.

\section{
Conclusion
\label{sec:conclusion}
}
In conclusion, we have presented the numerical integration approaches implemented to reduce by a factor of 90 the computational demand of the calculation of event probabilities using the ME technique. We achieve this by using low-discrepancy sequences for the numerical MC integration in conjunction with a dedicated estimator of the numerical uncertainty --- a novelty in the context of the ME technique, as well as the factorization of the jet energy scale factor $\kjes$ from the ME calculation --- newly applied in the context of \mt measurements with an {\it in situ} jet energy scale calibration. These improvements have been validated through MC studies.
The low-discrepancy sequences are universally applicable for numerical MC integration, and are not specific to the presented studies.

\section*{Acknowledments}
We thank our D0 colleagues for useful discussions and for their kind
permission to use the D0 detector simulation and other collaborative software
to expedite the preparation of this paper. The authors acknowledge the
support from 
the Department of Energy (USA), 
the National Science Foundation (USA),
the Bundesministerium f\"ur Bildung und Forschung (Germany), 
and the Deutsche Forschungsgemeinschaft (Germany).

\bibliographystyle{model1-num-names}
\bibliography{mtop_me_accelrn}

\end{document}